\documentclass[11pt,a4paper]{article}
\usepackage{graphicx}
\usepackage[latin1]{inputenc}
\usepackage[T1]{fontenc}
\usepackage{amsmath}
\usepackage{upgreek}
\usepackage[dvips]{epsfig}
\begin{document}

% ================================================================
%   Title
% ================================================================

\begin{titlepage} 
  \title{On the quantum mechanical description of \\ 
the interaction between particle and detector}
\author{Klaus Wick \\ \small{Institut f\"ur Experimentalphysik, University of Hamburg,} \\ \small{Hamburg, Germany}}
\thanks{e-mail: klaus.wick@desy.de}

\end{titlepage} 
\date{}
\maketitle

\begin{abstract}
Quantum measurements of physical quantities are usually described as ideal measurements. However, only a few measurements fulfil the conditions of ideal measurements.  The aim of the present work is to describe real position measurements with detectors that are able to detect single particles.  
For this purpose, a detector model has been developed that can describe the time dependence of interactions between nonrelativistic particles and a detector.  At the beginning of a position measurement, the detector behaves as a target consisting of a large number of quantum mechanical systems.  
The incident object interacts with a single atom, electron or nucleus, but not with the whole detector.  This reaction is a quantum mechanical process.  At the end of the measurement, the detector can be considered as a classical apparatus.  
A detector is neither a quantum mechanical system nor a classical apparatus. 
The detector model explains why one obtains a well-defined result for each individual measurement.  It further explains that, in general, it is impossible to predict the outcome of the next measurement.  
The main advantage is that it describes real rather than only ideal measurements. 
\end{abstract}

\emph{Keywords:} Quantum measurement process; Position measurement; State reduction  

% ================================================================
%   Introduction
% ================================================================

\section{Introduction} 

A measurement device may consist of various components (magnets, cavities, crystals, detectors and so on).  
In the literature, such a complex device is often referred to as a detector.  
In the following, the word detector will only be employed for the components of the measurement device that can provide an output signal.  
Examples include ionisation chambers, semiconducting detectors, photomultiplier tubes, scintillation counters, and cloud and bubble chambers.  
The output signal delivered by a detector indicates that a quantum object (particle or photon) has been detected.  
The detection of an object is also a position measurement, because at the moment of the measurement each detector has a well-defined position.  The uncertainty of the measured position is defined by the size of the detector.  
A measurement is complete when the detector has produced the output signal and the result has been registered.  Detectors are highly important for experimental performance, because measurement results for single particles or photons are usually determined by means of detectors.  

One problem is that quantum theory describes the interactions between quantum objects and detectors in a highly simplified manner.  
The Copenhagen interpretation of quantum mechanics assumes that a detector is a classical apparatus.  However, it is not clear how a quantum mechanical system can interact with a classical one.  
On the other hand, John von Neumann~\cite{Ne32} considers a detector as a quantum mechanical system.  An ideal measurement then describes an interaction between two quantum mechanical systems: a microscopic one (the quantum object) and macroscopic one (the detector\footnote{W.~H. Zurek~\cite{Zu91} discusses the question of whether macroscopic objects can be treated as quantum mechanical systems.}).  
Unfortunately, the time evolution of the wave functions of both systems during the measurement process cannot be described by the time-dependent Schr\"odinger equation.\footnote{A more detailed description of the so-called measurement problem of quantum mechanics is provided by M. Schlosshauer~\cite{Schl}.}   

Following an ideal measurement, the incident object will be in a well-defined quantum mechanical state.  Hence, an ideal measurement is repeatable, and the same object can interact a second time with an identical measurement device.  
If one repeats the same measurement \emph{immediately}, then one must obtain the same result as for the first measurement.  According to R.~Omn\`es~\cite{Om94}, this statement can be considered as the definition of an ideal measurement.  

Most real measurements do not fulfil the conditions of ideal measurements.  
Either they cannot be repeated, or if they are repeatable one will not obtain the same outcome as in the first measurement.  
Neutrons and photons are often absorbed when they strike a detector.  
If one wants to measure the energy of a charged particle, this particle has to deposit its full energy in the detector.  
Hence, these measurements are not repeatable.  

The first aim of the present work is to describe real position measurements.  
A necessary prerequisite for achieving this goal is a better understanding of the interaction between a quantum object and detector.  As a first step, the time dependence of this interaction will be studied in the next section.   

% ================================================================
%   The interaction of quantum object and detector
% ================================================================

\section{The interaction \\ between a quantum object and detector}\label{ch_qo-det} 

\begin{figure}[ht]                                      
\centering
\includegraphics[width=11.5cm]{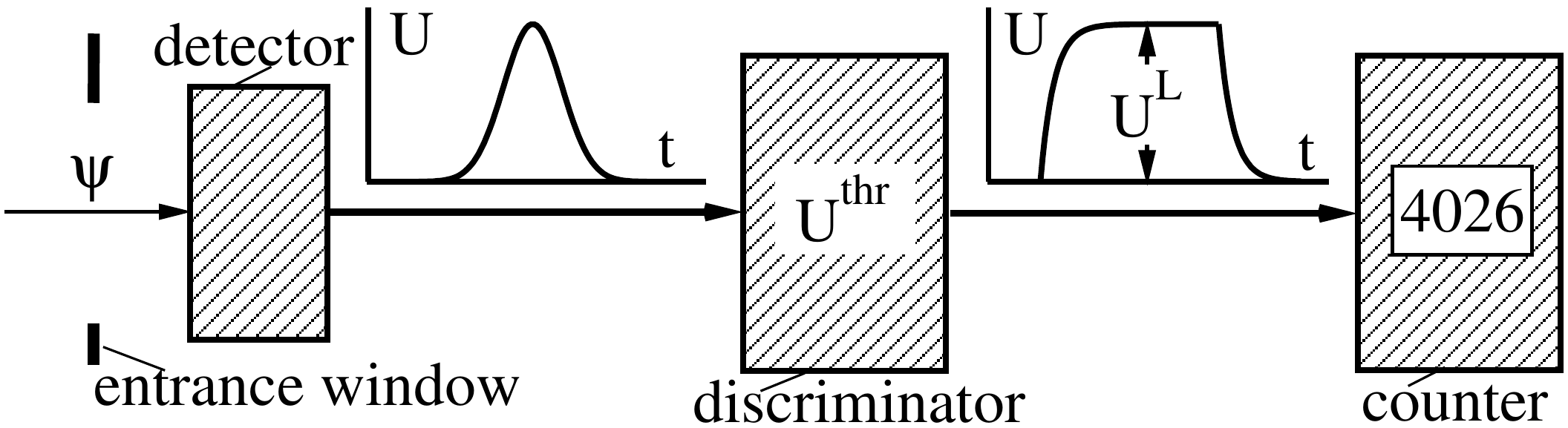}
\caption{Detection of a single particle or photon.  The voltage pulse $U(t)$ generated by the detector is transformed by the discriminator into a logical output signal $U^{\rm L}$ if $U(t)$ exceeds the threshold $U^{\rm thr}$.  
The output signal $U^{\rm L}$ increases the content of the counter memory by one.}\label{rAbb_pos-meas}   
\end{figure}

Figure \ref{rAbb_pos-meas} shows a simple electronic set-up that can detect a single quantum object.  An incident charged particle deposits a certain amount of energy in the detector.  
This energy is utilised to provide a voltage pulse $U(t)$ (where $t$ denotes time).  
To suppress thermal noise, the experimenter will set a fixed threshold $U^{\rm thr}$. The discriminator in Fig. \ref{rAbb_pos-meas} will only deliver a logical output signal $U^{\rm L}$ if $U(t)$ exceeds the threshold.  
In most detectors, this means that an output signal $U^{\rm L}$ can only be generated if the energy $E^{\rm dep}$ deposited in the detector is greater than a certain threshold energy $E^{\rm thr}$:  
\begin{equation}\label{r_E-thr}   
E^{\rm dep} > E^{\rm thr}  \;.
\end{equation} 

An incident charged particle (a) with sufficient energy can ionise and excite many atoms (X) during its passage through matter.  
\begin{eqnarray}
\mbox{a + X} & \to & \mbox{a + X}^* \;,  \label{r_Anrg}\\
\mbox{a + X} & \to & \mbox{a + X}^+ + \mbox{e}^- \;.   \label{r_Ionis}
\end{eqnarray}
In a semiconducting detector, a large number of electrons are transferred from the valence band to the conduction band.  
In a scintillator, many photons are produced.  
These examples show that the numbers of charged objects and photons are vastly increased along the path of a charged particle.  
Thus, a single particle state becomes a many particle state.  
In the following, this effect will be called amplification.  
Such amplification processes are employed in various detectors to produce a detectable output signal $U^{\rm L}$.  

Neutrons and photons do not ionise or excite atoms during their passage through matter.  Hence, one utilises special detectors to detect such objects.  
In these detectors, reactions in which one or two charged particles are released can occur.  
Examples include nuclear reactions for neutrons, and pair production and the photoelectric and Compton effects for photons.  
In the following, these reactions will be called start reactions, because the charged particles ionise and excite many atoms and initiate the amplification process mentioned above.  If condition (\ref{r_E-thr}) is fulfilled, then a logical output signal $U^{\rm L}$ will be produced.  

In the next step, we consider the time dependence of the interaction between a quantum object and detector.  
Successful measurements can often be divided into three phases:  

In \emph{phase~1}, the start reaction occurs.  
The incident object interacts with a microscopic part of the detector (atom, molecule, electron or nucleus), rather than with the whole detector.  

\emph{Phase~2} is the amplification phase. An avalanche of secondary objects (charged particles and/or photons) is released.  

\emph{Phase~3} is the readout phase. A fully operable detector will produce a detectable output signal.  

As a first example, we consider the detection of single photons of visible light with a photomultiplier tube.  The photomultiplier tube consists of three parts: a photocathode, dynode system, and anode.  
In phase~1, the incident photon interacts with a single atom in the photocathode.  This is the start reaction.  
The photon will be absorbed, and a photoelectron will be ejected.  
In this case, the photoelectron will not fulfil condition (\ref{r_E-thr}).  
Hence, it will be accelerated to the first dynode in an electric field.  
Here, it will eject several secondary electrons, which are accelerated to the second dynode.  
In phase~2, many secondary electrons are released in the dynode system.  In phase~3, a voltage pulse $U(t)$ is readout at the anode, and the discriminator in Fig. \ref{rAbb_pos-meas} will produce a detectable output signal $U^{\rm L}$.  

Thermal neutrons can be detected in a proportional chamber filled with boron trifluoride (BF$_3$) gas enriched to 96\% $^{10}$B.  Here, the reaction 
\begin{equation}\label{r_Bor-Fluo} 
\mbox{n} + \mbox{$^{10}$B} \to \mbox{$^{11}$B}^* \to \alpha + \mbox{$^7$Li}  
\end{equation}
is the start reaction.  
The charged particles $\alpha$ and $^7$Li ionise many molecules in phase~2.  
Electrons and ions move to the electrodes of the chamber in an external electric field and produce the output signal.  
Both examples show that the incident object (photon or neutron) interacts with an atom or nucleus in phase~1, but not with the whole detector.  

Many reactions occur in phases~1 and 2 of a position measurement.  
Although each reaction is a quantum mechanical process, the sum effect of these reactions (the production of incoherent light or an electric current in phase~3) can be described in the language of classical physics.  
A scintillation counter collects the light emitted by many excited atoms at the photocathode of a photomultiplier tube.   
In a semiconducting detector, electrons and holes are accelerated in an external electric field and produce a current pulse.  
 
A start reaction initiates the amplification process in a detector, and with certainty produces a logical output signal $U^{\rm L}$.  
Neutrons and photons can produce various reactions that do not start an amplification process in the detector.  
Hence, we define the start reaction as the first interaction occurring in a detector that fulfils the following two conditions:  

(I) A start reaction is an interaction between two quantum objects.  

(II) In the exit channel of a start reaction, there must be at least one or two charged particles that are able to start the amplification process and provide the logical output signal $U^{\rm L}$.  

When charged particles interact with a detector that is placed under vacuum conditions, we define its first interaction with an atom of the detector as the start reaction.  The first interaction may be an elastic or inelastic process.  
This definition fulfils conditions (I) and (II).  

An experimenter who wants to detect particles will first choose an appropriate detector, and then bring it into an operable state.  In some cases, there only exists one start reaction. Reaction (\ref{r_Bor-Fluo}) is an example of this.  
On the other side, an incident photon can initiate various reactions that fulfil the conditions (I) and (II) (pair production, and the photoelectric and Compton effects).  In the following, we define the `start reaction' as consisting of all reactions fulfilling both conditions (I) and (II).  
A particle that produces a start reaction in a detector will produce an output signal with certainty.  

% ================================================================
%   Detector model for position measurements
% ================================================================

\section{Detector model for position measurements}\label{ch_det-mod} 

Interactions between quantum objects and different types of detectors can be considered as sequences consisting of three phases (start reaction, amplification, and readout).  It will be shown in~\cite{Wick} that this statement holds for neutrons, photons, atoms, and nonrelativistic charged particles.  
This observation will be utilised to develop a detector model that can describe individual measurements of quantum objects.  
In the following, this model will be called the three-phase model.  

Only detectors that fulfil the following requirements will be considered:  

(A) The detector should be able to detect single particles or photons.  

(B) The above-mentioned amplification process should be initiated by reactions of type (\ref{r_Anrg}) or (\ref{r_Ionis}).  

These requirements are fulfilled when photons, neutrons or nonrelativistic charged particles interact with different types of detectors.  
Examples include ionisation chambers, semiconducting detectors, photomultiplier tubes, and scintillation counters.\footnote{The photographic process does not fulfil requirement (A).  Detectors that use the Cherenkov effect do not fulfil requirement (B).} In all cases, the quantum mechanical state of the incident object is destroyed when it is detected in a detector.  

If the incident object performs a start reaction in a fully operable detector, then the same detector will provide a logical output signal $U^{\rm L}$ with certainty. For detectors that fulfil the requirements (A) and (B), the converse is also true.  
If the detector delivers a logical output signal, then an amplification process must have occurred that was initiated by a start reaction in the same detector.  
We conclude that there is a one-to-one correlation between a quantum mechanical event (a start reaction occurs in phase~1) and a classical event (the detector and discriminator produce a logical output signal $U^{\rm L}$ in phase~3).  

The reactions occurring in phase~2 appear to have no influence.  
However, this is not true.  They are important if one measures the energies of charged particles or if one investigates tracks of particles in a bubble chamber.  
However, in a position measurement one is only interested in the question of whether a logical output signal has been produced.  This yes/no decision depends only on the question of whether a start reaction has been performed.
In the following, only position measurements will be discussed.  

So far, we have only studied the interactions between quantum objects and a single detector.  In the following example, we will consider interactions between charged particles and an array of several ($Z$) small ionisation chambers.  
This may be useful if one wants to measure the angular distribution of reaction products of a certain reaction.  
From the classical viewpoint, an ionisation chamber is a capacitor filled with argon gas.  From the quantum mechanical viewpoint, the detector is a collection of atoms or molecules.  
In phase~1, the state of the projectile is described by a probability wave that can interact with all atoms of all detectors.  Each incident particle views the different detectors as one big target consisting of all the atoms of all the detectors.  
The spatial order of the detectors does not play an essential role in phase~1 of the measurement.  

The wave function of the projectile collapses at the moment at which it interacts with a single atom of a detector.  Let us assume that the start reaction randomly occurs in detector D$_m$ (with $1 \le m \le Z$).  
Because the mean free path length of a charged particle is considerably smaller than the size of a typical detector, all the reactions following in phase~2 will occur in the same detector D$_m$. At this point, the spatial order of the detectors is important.  
In phase~3, the ionisation chamber D$_m$ can be considered as a capacitor in which clouds of positively and negatively charged particles move.  
Under the influence of an external electric field, a large number of electrons and ions move towards the capacitor plates.  
The uncorrelated movement of several tens of thousands of charged particles can be described as a classical electric current.  
This current generates a short voltage pulse in detector D$_m$.  
Before measurement, the outputs of all discriminators were in the ground state ($U_n^{\rm L}=0$~V for $n=1~- Z$). Immediately after the detection of the particle, for a short time one detector (D$_m$) will be in an `excited' state with a definite output voltage ($U_m^{\rm L}=1$~V).    
\[ U_n^{\rm L} = \left\{ \begin{array}{rl} 1\,  \mbox{V for} & n=m \;, \\
0\,  \mbox{V for} &  n \ne m  \;. \end{array} \right\}    \]
Only detector D$_m$ will produce a logical output signal.  
This indicates that the particle has been detected in detector D$_m$.  
A well-defined outcome will be obtained for each successful measurement.  
However, one cannot predict the detector in which the incident particle will be detected, because it is impossible to predict the atom with which it will produce a start reaction.  

% ================================================================
%   Probability distribution of measurement results
% ================================================================

\section{Probability distribution \\ of measurement results}\label{ch_prob-dis}

In the present work, we are only interested in measurements of physical quantities for which quantum theory only provides statistical predictions.  
Again, we consider position measurements with $Z$ small detectors D$_n$.  
The aim of the investigation is to determine the probability distributions of measurement results.  These distributions are determined in quite a different manner in experiment and theory.  
For the experimenter, the detector and discriminator are classical apparatuses, which produce a yes/no decision for each object that leaves the source. The discriminator will either generate a logical output signal or deliver no signal.  This yes/no decision is described in the language of classical physics.  

Let us assume that the source has emitted $N^{\rm s}$ particles during an experiment, whose state is described by the same wave function, and 
$N({\rm D}_n)$ particles have been detected by detector D$_n$.  
For each detector, the experimenter determines the probability $\Delta P({\rm D}_n)$ that detector D$_n$ and the corresponding discriminator (see Fig. \ref{rAbb_pos-meas}) generate a logical output signal $U_n^{\rm L}$.  
In the limit $N^{\rm s} \to \infty$, the probabilities $\Delta P({\rm D}_n)$  are defined as the ratios $N({\rm D}_n)/N^{\rm s}$ (with $n = 1 - Z$).  

The (classical) probability $\Delta P({\rm D}_n)$ cannot be calculated in the framework of quantum mechanics.  However, the probability $\Delta P(\mbox{\boldmath$R$}_n)$ that a nonrelativistic charged particle has passed the entrance window of detector D$_n$ (see Fig. \ref{rAbb_pos-meas}) can be determined from the wave function of the particle.  The centre of this entrance window is located at position $\mbox{\boldmath$R$}_n$.  
For single events, one can of course obtain only statistical predictions.  
On average, $N_n = N^{\rm s} \Delta P(\mbox{\boldmath$R$}_n)$ particles will pass this entrance window.   
If these charged particles fulfil condition (\ref{r_E-thr}), then they will initiate a start reaction in detector D$_n$ with certainty, and produce a detectable output signal $U_n^{\rm L}$.  Hence, the number $N_n$ must be equal to the number of detected particles $N({\rm D}_n)$.  
For incident charged particles, one can conclude that the probabilities $\Delta P(\mbox{\boldmath$R$}_n)$ and $\Delta P({\rm D}_n)$ are equal:  
\begin{equation}\label{r_pos-m-Dn}
\Delta P(\mbox{\boldmath$R$}_n) = \frac{N_n}{N^{\rm s}} 
= \frac{N({\rm D}_n)}{N^{\rm s}} = \Delta P({\rm D}_n) 
\quad \mbox{for}  \quad n = 1, \, 2,\dots,Z \;.  
\end{equation}
It should be noted that $\Delta P({\rm D}_n)$ is determined by the experimenter using methods of classical physics, while $\Delta P(\mbox{\boldmath$R$}_n)$ is calculated in the framework of quantum mechanics.  From (\ref{r_pos-m-Dn}), one can conclude that the measured probability distribution $\Delta P({\rm D}_n)$ can be reproduced using the methods of quantum mechanics.  

The last statement also holds for neutrons, as will be shown in~\cite{Wick}.  In this case, the probability $\Delta P_n^{\rm St}$ that the incident object performs a start reaction in detector D$_n$ can be calculated using the methods of quantum physics.  
One can show that $\Delta P_n^{\rm St}$ is equal to the measured probability $\Delta P({\rm D}_n)$.  Here, we use the statement of the three-phase model (see Section \ref{ch_det-mod}) that there is a one-to-one correlation between the start reaction (one of the possible start reactions) in detector D$_n$ and the production of a logical output signal $U_n^{\rm L}$.  

% ================================================================
%   Summary and outlook
% ================================================================

\section{Summary and outlook}   

In the preceding sections, we have mainly studied interactions between quantum objects and a small number of detectors.  
The incident object is in a quantum mechanical state.  
Conversely, the detector and discriminator produce a yes/no decision, and provide classical information.  

For various types of detectors (such as ionisation chambers, semiconducting detectors, photomultiplier tubes, and scintillation and proportional counters) the measurement process can be considered as a sequence of three phases (the start reaction, amplification, and readout).  
This observation has been utilised to develop the three-phase model, which can describe individual measurements for nonrelativistic particles.  

Detectors behave quite differently during the three phases.  Let us assume that charged particles strike an array of $Z$ ionisation chambers.  
In phase~1, the $Z$ detectors form one big target with a large number of quantum mechanical systems.  
In the start reaction, the incident particle interacts with one of these systems (atom, molecule, nucleus, or electron), but not with the whole detector.  
If the start reaction occurs in detector D$_1$, then all reactions following in phase~2 will occur in the same detector.  
All reactions are quantum mechanical processes.  
In phases~1 and 2, the incident particle ionises a large number of atoms and produces a cloud of charged particles in detector D$_1$.  In phase~3, this cloud determines the properties of the output signal, while the incident particle does not play a role.  
At the beginning of a position measurement, the detector can be considered as a collection of many quantum mechanical systems (atoms), and at the end the detector behaves like a classical apparatus.  
We conclude that the detector is neither a quantum mechanical system nor a classical apparatus.  The transition from the quantum mechanical to the classical description occurs in the detector.  

Probability distributions are determined quite differently in experimental and theoretical physics.  The experimenter employs methods of classical physics to determine the probability that detector D$_n$ generates an output signal.  
This probability cannot be calculated in the framework of quantum physics.  
However, one can calculate the probability that the incident object performs a start reaction in detector D$_n$.  
According to the three-phase model, both probabilities are equal.  
Here, the one-to-one correlation between the start reaction and logical output signal $U_n^{\rm L}$ is utilised.  

For position measurements with detectors that fulfil the requirements (A) and (B) (in Section \ref{ch_det-mod}), the essential properties of the three-phase model can be summarised as follows:  
\begin{itemize} 
\item[-] The model explains how nonrelativistic particles can interact with a detector in the framework of quantum physics.  
\item[-] It explains why it is impossible to predict the outcome of the next measurement.  The outcome will be determined at random, because one cannot predict the detector in which the incident particle will perform a start reaction.  
\item[-]  If the start reaction occurs in detector D$_n$, the same detector will provide the classical information comprising the logical output signal $U_n^{\rm L}$.  This signal indicates the measurement result, comprising the position of the detected particle. 
\end{itemize}

Position measurements play an important role in many other measurements.  
This is because of the fact that position coordinates can be measured directly, unlike practically all other physical quantities.  
As an example, let us consider the measurement of the spin component $S_z$ in the Stern--Gerlach experiment.  Each measurement is a two-step process.  
In the first step, the incident particle (atom) interacts with an inhomogeneous magnetic field, and in the second step a position measurement is performed.  
A particle that has passed the magnetic field is in an entangled state.  
The spin state and the centre-of-mass motion of the particle are correlated (see~\cite{Pa58} and \cite{GoYa}).  The spin component $S_z$ of the particle cannot be measured directly, but it can be determined by a position measurement with a correctly calibrated Stern--Gerlach apparatus.  
The detection of the particle in one of the two detectors leads to a state reduction.  At the moment of the measurement, the particle is in a state with a well-defined spin component $S_z$.  

The three-phase model is extended in such a manner that it can describe the interaction between the incident particle and the whole measuring device.  
In phase~1, the particle interacts with the magnetic field and performs a start reaction in one of the two detectors.  
Both processes can be described in the framework of quantum physics.  
The amplification process in phase~2 and the readout in phase~3 are similar to the corresponding processes in a position measurement.  

A more general discussion of quantum measurements will be provided in~\cite{Wick}.  In addition, the question of whether one can find a solution for the measurement problem of quantum mechanics will be discussed.  \\

% ================================================================
%   Acknowledgments
% ================================================================

\section*{Acknowledgments}

For helpful discussions, I am grateful to Prof. Dr. J.~Bar\-tels, 
Prof. Dr. K.~Fre\-den\-hagen, Prof. Dr. R.~Klan\-ner and 
Prof. Dr. E.~Lohr\-mann of the University of Hamburg.  

% ================================================================
%   References
% ================================================================

\end{document}